\documentclass[3p,twocolumn,sort&compress]{elsarticle}

\usepackage{amsmath,amssymb}
\usepackage[colorlinks]{hyperref}
\usepackage{ulem}
\usepackage{xcolor}

\usepackage{cleveref}

\renewcommand{\t}[1]{\mathrm{{#1}}}

\journal{Physics Letters A}

\begin{document}

\begin{frontmatter}

\title{Evidence for structural damping in a high-stress silicon nitride nanobeam and its implications for quantum optomechanics}

\author[EPFLaddress]{S. A. Fedorov}
\author[EPFLaddress]{V. Sudhir}
\author[EPFLaddress]{R. Schilling}
\author[EPFLaddress]{H. Sch\"{u}tz}
\author[EPFLaddress]{D. J. Wilson}
\author[EPFLaddress]{T. J. Kippenberg\fnref{corrfnt}}
\address[EPFLaddress]{Institute of Physics, {\'E}cole Polytechnique F{\'e}d{\'e}rale de Lausanne, 
	Lausanne 1015, Switzerland}
\fntext[corrfnt]{E-mail: tobias.kippenberg@epfl.ch.}

\begin{abstract} 
We resolve the thermal motion of a high-stress silicon nitride nanobeam at frequencies far below its fundamental flexural resonance (3.4 MHz) using cavity-enhanced optical interferometry. Over two decades, the displacement spectrum is well-modeled by that of a damped harmonic oscillator driven by a $1/f$ thermal force, suggesting that the loss angle of the beam material is frequency-independent. The inferred loss angle at 3.4 MHz, $\phi = 4.5\cdot 10^{-6}$, agrees well with the quality factor ($Q$) of the fundamental beam mode ($\phi = Q^{-1}$).
In conjunction with $Q$ measurements made on higher order flexural modes, and accounting for the mode dependence of stress-induced loss dilution, we find that the intrinsic (undiluted) loss angle of the beam changes by less than a factor of 2 between 50 kHz and 50 MHz. We discuss the impact of such ``structural damping" on experiments in quantum optomechanics, in which the thermal force acting on a mechanical oscillator coupled to an optical cavity is overwhelmed by radiation pressure shot noise. As an illustration, we show that structural damping reduces the bandwidth of ponderomotive squeezing.
\end{abstract}


\end{frontmatter}

\section{Introduction}

Nanomechanical oscillators are widely used as resonant force detectors, achieving sensitivities as low as aN/$\sqrt{\t{Hz}}$ by dint of their low mass and stiffness \cite{binnig_1986,rugar_mechanical_1992}. The force sensitivity of a nanomechanical oscillator is fundamentally limited by its thermal motion, which, over a narrow band, is well-approximated as the response to a Langevin force with a white noise spectrum \cite{saulson_thermal_1990,spectraldensity},
\begin{equation}\label{SFth}
S_{F}^\t{th}[\Omega] = 4k_B T m \Gamma_\t{m}.
\end{equation}
According to eq. \eqref{SFth}, force sensitivity can be improved by reducing the mass ($m$), damping rate ($\Gamma_\t{m}$), and temperature ($T$) of the oscillator.  For an oscillator with frequency $\Omega_\t{m}$, the second factor corresponds to increasing the oscillator's quality factor, $Q=\Omega_\t{m}/\Gamma_\t{m}$. Advances in nanomechanics have largely followed efforts to increase $Q/m$ \cite{imboden2014dissipation}.  

In recent years, integration of high-$Q$ nanomechanical resonators with high finesse optical cavities \cite{aspelmeyer_cavity_2015} has enabled thermal-noise-limited force measurements far from resonance \cite{wilson2015measurement}, as well as observation of radiation pressure shot noise, an example of quantum measurement back-action \cite{purdy_observation_2013,teufel2016overwhelming}.  These advances open the door to a new class of broadband, high sensitivity nano-optomechanical force sensors \cite{norte2016mechanical,reinhardt2016ultralow}.  They also offer the opportunity to study the quantum limits of force and displacement measurement \cite{braginsky1995quantum,braginsky_weak_1977}, a subject with a rich history in the context of gravitational wave detection \cite{corbitt2004review}. 

For broadband force sensing using mechanical oscillators, an important consideration is the color (frequency-dependence) of the thermal force, which depends on that of the damping mechanism.  In general the thermal force may be expressed as \cite{saulson_thermal_1990}
\begin{equation}\label{SFth_general}
S_{F}^\t{th}[\Omega] = -\frac{4k_B T}{\Omega} \cdot\t{Im} \left(\chi [\Omega]^{-1}\right)
\end{equation}
where 
\begin{equation}
\chi[\Omega] = \frac{m^{-1}}{\Omega_\t{m}^2(1-i\phi[\Omega])-\Omega^2}
\end{equation}
is the oscillator susceptibility and $\phi[\Omega]$ is a generally frequency-dependent loss angle.  White thermal noise corresponds to a linear loss dispersion $\phi[\Omega]= Q^{-1}\cdot\Omega/\Omega_\t{m}$, and arises for viscous (velocity-proportional) damping mechanisms such as gas damping \cite{verbridge2008size}. By contrast, in the absence of external loss, internal modes of a solid-state mechanical resonator typically exhibit a frequency-independent loss angle, $\phi[\Omega]=Q^{-1}$ \cite{saulson_thermal_1990} --- so-called "structural damping".  The resulting pink ($1/f$, where $f=\Omega/2\pi$) thermal force fundamentally limits force sensitivity at frequencies below the resonance.  It also impacts the bandwidth of correlations produced by quantum-limited measurements, such as ponderomotive squeezing \cite{sudhir2017appearance} in the context of cavity optomechanics \cite{aspelmeyer_cavity_2015}. 
 
Though well-studied for precision macroscopic mechanical oscillators such as pendulum springs \cite{gonzalez_brownian_1995,bernardini_1999,yang_2009,neben2012structural}, little direct evidence of structural damping has been presented for nanomechanical oscillators. The challenge is two-fold:  First, it is difficult to realize nanomechanical oscillators which are limited by internal loss.  Second, it is difficult to resolve the thermal motion ($x$) of a nanomechanical oscillator far below resonance.  To wit
\begin{subequations}\label{eq:Sx}\begin{align}
S_{x}^\t{th}[\Omega]&=|\chi[\Omega]|^2 S_{F}^\t{th}[\Omega]\\ &\xrightarrow{\Omega\ll\Omega_\t{m}}S_{x}^\t{th}[\Omega_\t{m}]\cdot\phi[\Omega_\t{m}]\phi[\Omega]\cdot\frac{\Omega_\t{m}}{\Omega}
\end{align}\end{subequations}
which for a typical internal loss angle of $\phi=10^{-4}$ implies a 70 dB reduction of thermal motion at frequencies one octave below resonance ($\Omega/\Omega_\t{m}<0.1$).\par 
 
Here we study the broadband thermal motion of a high-stress silicon nitride (SiN) nanobeam, employing a microcavity-enhanced optical interferometer to achieve high sensitivity at frequencies two decades below the beam’s fundamental flexural resonance ($3.4$ MHz). The high signal-to-noise ratio over the large bandwidth allows us to directly distinguish between velocity and structural damping. In our measurements, we directly witness structural damping as a $1/f$ displacement noise spectrum from 50 kHz to 3.5 MHz. The calibrated magnitude of the noise is consistent with a loss angle of $\phi_0=(3\pm 1.2)\cdot 10^{-6}$, which agrees well with the value inferred from the near-resonant response of the fundamental mode. Notably, $\phi_0$ is 80 times smaller than the typical loss angle of bulk SiN \cite{villanueva_surface_loss_2014}.  This is because stress increases the elastic energy stored in the beam, resulting in ``dilution" of the intrinsic (unstressed) loss coefficient \cite{gonzalez1994brownian,Kotthaus_damping_PRL_2010}. Taking into account loss dilution and the measured $Q$ of higher order flexural modes, we infer that the intrinsic loss angle of the beam material is nearly constant, $(2.5\pm 1.5)\cdot10^{-4}$, between 50 kHz and 50 MHz. This finding is consistent with extensive measurements of $Q$ factors of $\mathrm{Si_3N_4}$ beams and membranes \cite{Kotthaus_damping_PRL_2010,chakram_dissipation_2014,yuRegal_control_damping_2012,villanueva_surface_loss_2014,ghadimi_dissipation_2016}, and suggests that surface loss dominates at our beam thickness \cite{villanueva_surface_loss_2014}. 

\begin{figure}[t!]
	\centering
	\includegraphics[width=\columnwidth]{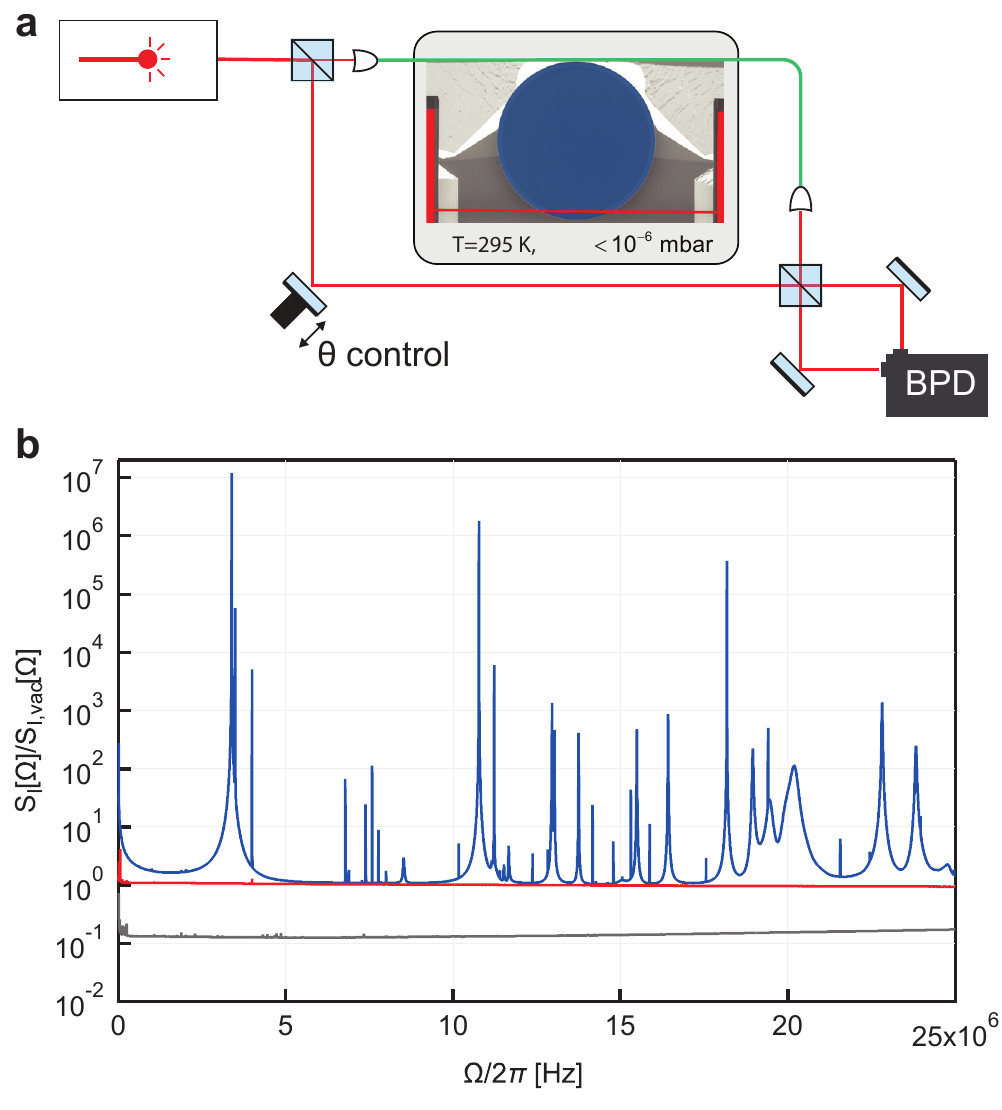}
	\caption{\label{fig:setup}
		(a) Schematic of the experiment. Motion of a nanobeam (red) is detected by coupling it to an optical
		microdisk cavity (blue) embedded in a balanced homodyne interferometer.   Light is coupled into the cavity using tapered fiber (green). The relative phase ($\theta$) between the two arms of the 
		interferometer (signal and LO) is stabilized using a piezo-actuated mirror. Nanobeam, microdisk, and tapered fiber are enclosed in a vacuum chamber.
		(b) Spectrum of the homodyne photocurrent $S_I[\Omega]$ (blue) when operating on the phase quadrature ($\theta=\pi/2$), normalized to the frequency average of the shot noise spectrum $S_{I,\t{vac}}[\Omega]$ (red).  $S_{I,\t{vac}}[\Omega]$ is obtained by blocking the signal beam and subtracting detector noise (gray curve, obtained by blocking both signal and LO beams).  Resonant motion of the fundamental beam mode is visible in $S_I[\Omega]$ in at $\Omega=2\pi\cdot3.4$ MHz.
	}
\end{figure}

In the following section we describe the experiment and measurement results in detail.  In Section \ref{sec:discussion} we discuss the impact of structural damping on the performance of cavity optomechanical systems.  We consider specifically a recently accessed regime in which the thermal force of a nanomechanical oscillator is overwhelmed by radiation pressure shot noise, giving rise to ponderomotive squeezing of the cavity light field.  We show that structural damping in this case reduces the bandwidth over which squeezing can be observed.

\section{Experimental details and results}\label{sec:experiment}

\begin{figure*}[t!]
	\centering
	\includegraphics[width=\textwidth]{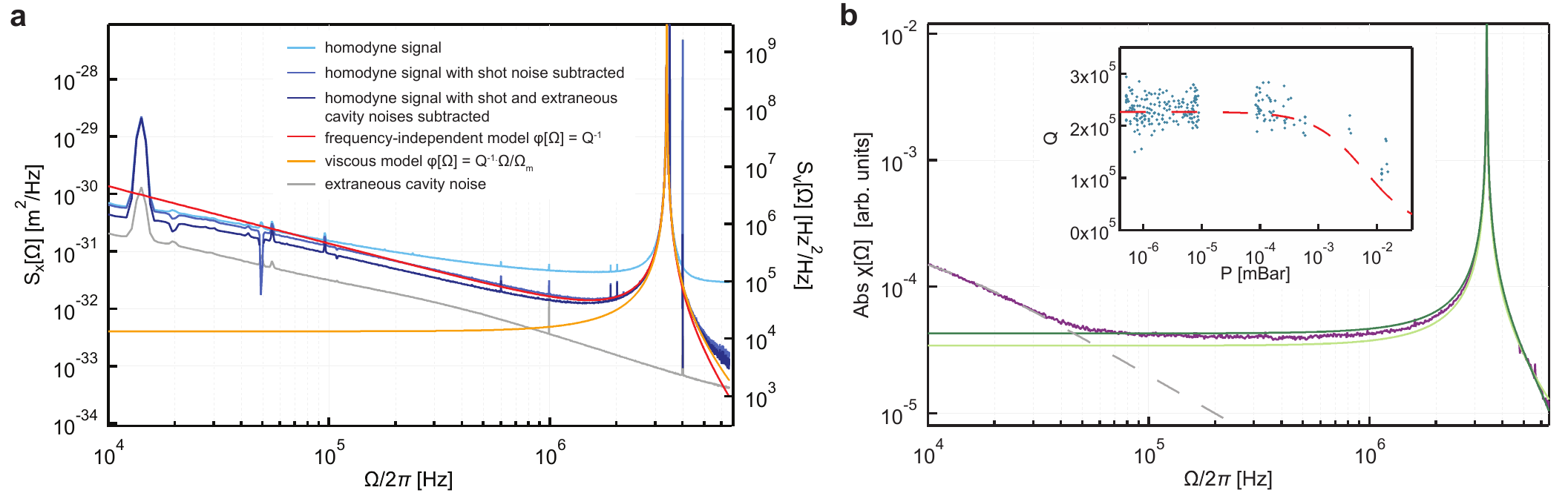}
	\caption{\label{fig:thermNoise}
	(a) Broadband thermal motion of a Si$_3$N$_4$ nanobeam measured using a microcavity-enhanced optical interferometer. Light blue curve is the unprocessed homodyne signal spectrum. Blue and dark blue curves correspond to, respectively, spectrum after subtraction of the local oscillator and after subtraction of the estimated extraneous cavity noise (gray). Red (orange) curve is a structural (velocity) damping model (see main text for details).
	(b) Response of the cavity resonance frequency to a coherent radiation pressure force, used to estimate the susceptibility of the nanobeam, $\chi[\Omega]$. Measured response is shown in violet. Light and dark green curves are models of $|\chi[\Omega]|$ which take into account only the fundamental mode and modes $n=\{1,3,5\}$ of the nanobeam, respectively. Dashed gray curve is a fit of photothermal contribution. The inset shows measurements of $Q$ versus gas pressure obtained from the width of the thermal noise peak. Blue points are measurements.  Dashed red line is an analytical estimate taking into account both free space and squeeze-film gas damping \cite{martin_gas_damp_2008,bao_squeeze_2007}.} 
\end{figure*}

An overview of the experiment is given in \cref{fig:setup}.
The nanomechanical resonator used for our study is a 80-nm-thick, 200-nm-wide, 70-$\mu$m long beam made of high-stress ($\sigma\approx 0.8$ GPa) Si$_3$N$_4$.  To minimize gas damping (\cref{fig:thermNoise}b, inset), the beam is placed in a vacuum chamber at $5\cdot 10^{-7}$ mbar, resulting in an internal-loss-limited damping rate of $\Gamma_\t{m} = 2\pi\cdot 15$ Hz for its fundamental flexural mode ($\Omega_\t{m} = 2\pi\cdot 3.4$ MHz).
To read out its motion, the beam is dispersively coupled to a high finesse silica microdisk cavity embedded in one arm of an balanced homodyne interferometer \cite{anetsberger_near-field_2009,schilling_near-field_2016}. Light coupled resonantly through the cavity (using a tapered fiber) experiences a phase shift $\delta \omega_c(t)/\kappa$, where $\kappa$ is the cavity decay rate and $\delta\omega_c(t)=G x(t)$ is the cavity resonance frequency shift produced by out-of-plane deflection of the beam's midpoint $x(t)$.  The phase shift is enhanced by engineering the beam in the evanescent near-field of the microdisk \cite{schilling_near-field_2016}, resulting in a frequency pulling factor of $G = 2\pi\cdot 1.8$ GHz/nm and a critically coupled cavity decay rate of $\kappa = 2\pi\cdot 4.5$ GHz at an operating wavelength of $\lambda = 780$ nm.  
To convert the homodyne signal to cavity frequency noise units, a calibrated phase modulation tone is applied to the laser \cite{gorodetsky_determination_2010,schilling_near-field_2016}.

Thermal motion of the beam is observed in the output photocurrent of the homodyne detector. In general the photocurrent contains various noise components, which we distinguish using the sequence of measurements shown in \cref{fig:setup}b. For the data shown, a cavity input (signal) power of 24 $\mu$W and a local oscillator (LO) power of 1 mW was used, both provided by a tunable diode laser.  The relative phase ($\theta$) of signal and LO beams was stabilized using a piezo-actuated mirror.   Maximum sensitivity to the beam's motion is obtained on the phase quadrature, $\theta = \pi/2$ (blue curve). Thermal motion of the fundamental mode is in this case visible as a Lorentzian noise peak in the photocurrent spectrum at 3.4 MHz.  The imprecision (noise floor) of the measurement has contributions due to LO shot noise (red curve) and photodetector noise (gray curve), characterized by blocking the signal beam and both the signal and the LO beams, respectively.

To measure thermal motion of the nanobeam, shot noise is subtracted from the phase-quadrature signal spectrum in \cref{fig:setup}b. An estimate of extraneous cavity noise, including termorefractive noise and thermomechanical noise of the disk modes, inferred from measurements made on a bare microdisk is also subtracted (see discussion below).  The result is shown in blue in \cref{fig:thermNoise}a, in units of cavity frequency noise, $S_{\omega_\t{c}}[\Omega]$ (right axis), and displacement noise, $S_x[\Omega] = G^{-2}S_{\omega_\t{c}}[\Omega]$ (left axis).  $G$ is determined by normalizing the area underneath the fundamental noise peak to $\langle \omega_\t{c}^2 \rangle = G^2\cdot k_\t{B}T_\t{eff}/m\Omega_\t{m}^2$, where $T_\t{eff}$ is the effective mode temperature. (For this task, laser power is reduced until radiation pressure damping is negligible, so that $T_\t{eff}\approx 295$ K). The absolute magnitude of $G$ and $S_x[\Omega]$ depends on the definition of the effective mass, $m$.  For deflection of the midpoint of the beam, $m = m_\t{phys}/2$ where $m_\t{phys}$ is the beam's physical mass.  We assume the theoretical value $m_\t{phys}=1.9$ pg using a SiN density of 3100 kg/m$^3$ and taking into account a small defect in the center of the beam used to increase optomechanical coupling \cite{sudhir_RT_QBA_2016}.

The $1/f$ scaling of the apparent displacement noise spectrum at low frequencies suggests that the nanobeam is structurally damped.  To support this hypothesis, we consider three pieces of evidence: (1) absence of systematic error which might otherwise produce the observed $1/f$ tail, (2) validity of the model for the mechanical susceptibility (eq. \eqref{SFth_general}b) based on a driven response measurement, and (3) correspondence between the apparent displacement noise and the thermal force model (eq. \eqref{SFth_general}a).

\begin{figure}[t!]
	\centering
	\includegraphics[width=\columnwidth]{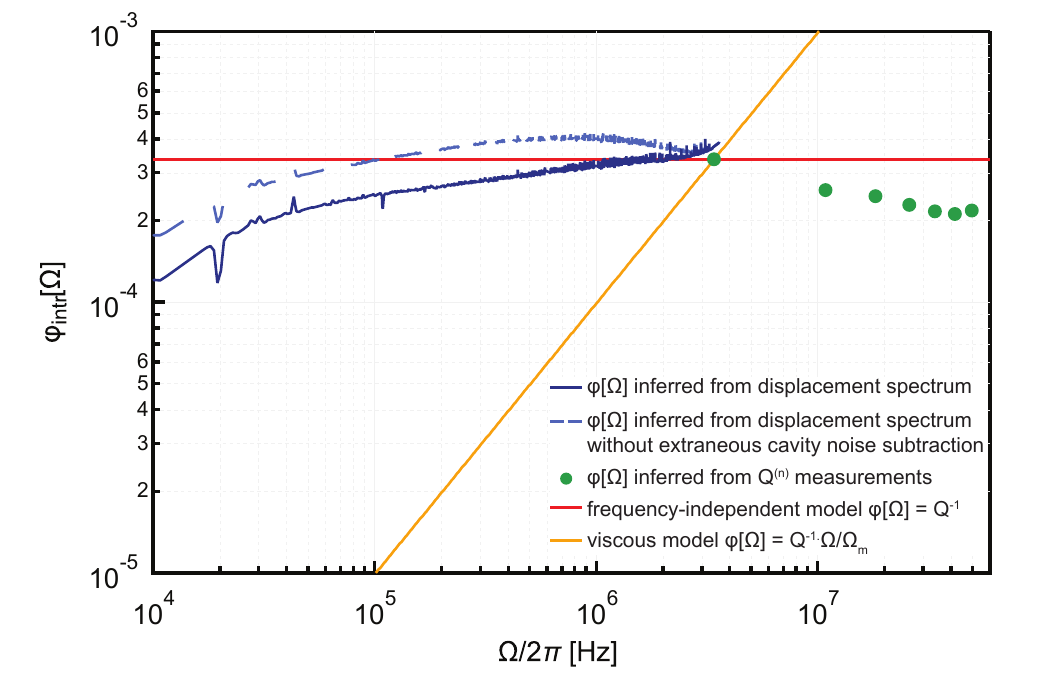}
	\caption{\label{fig:phiOmega}
		Intrinsic (undiluted) loss angle of a 80-nm-thick Si$_3$N$_4$ nanobeam inferred from its low frequency displacement spectrum (blue curve), the spectrum without cavity noise subtraction (light blue dashed curve) and linewidth measurements of various flexural modes (green dots). The red and orange solid curves show the loss angle variation, predicted by the frequency-independent loss angle model and viscous damping model, correspondingly. 
	}
\end{figure}

(1) Sources of systematic error which might produce the observed $1/f$ tail include non-uniform detector gain and extraneous low frequency noise. Non-uniform photodetector gain is inconsistent with the shot noise measurement, which is white within 15\% from 10 kHz to 7 MHz.  We therefore rule out this possibility. To estimate the contribution of extraneous noise, the frequency noise of a bare microdisk is measured. 
Within the entire measurement range the bare disk noise is at least 6 dB smaller than that of the beam-coupled microdisk, and, even at the highest frequency, at least 3 dB larger than the approximately constant $\approx 30\,\t{Hz}/\sqrt{\t{Hz}}$ diode laser frequency noise \cite{sudhir_appearance_2017,sudhir_RT_QBA_2016}. The magnitude and frequency dependence of this background is consistent with thermorefractive noise, a well-known limit to the stability of  whispering gallery mode resonators \cite{gorodetsky2004fundamental,anetsberger_measuring_2010}. As shown in \cref{fig:thermNoise} and \cref{fig:phiOmega}, subtracting this noise leads to up to 30\% reduction in the apparent loss angle. Finally, we note that thermal motion of high order beam modes also contributes to the low frequency noise spectrum.  In our system, only odd-ordered out-of-plane modes are coupled to the cavity.  Relative to the fundamental mode, their contribution at low frequency scales as $(1/n^3)$, where $n$ is the mode order. The two nearest  modes ($n=\{3,5\}$) contribute less than 12\%.

(2) To verify the form of the mechanical susceptibility in eq. \eqref{SFth_general}, we monitor the response of the nanobeam to a controlled external force.  The force is applied by injecting an auxiliary, amplitude-modulated laser field into the cavity at $\lambda = 850$ nm, resulting in a sinusoidal radiation pressure.  The magnitude of the radiation pressure force is held roughly constant while sweeping its frequency between 10 kHz and 7 MHz.  The apparent displacement response, shown in \cref{fig:thermNoise}b, agrees well with that of a damped harmonic oscillator (dark green trace) for frequencies above 50 kHz. Notably, the response function, in contrast to the thermal force, is insensitive to the frequency dependence of the loss angle as long as the loss is small ($\phi[\Omega]\ll 1$).  In particular, for $\Omega\ll \Omega_m$, the response should be nearly flat, i.e.
\begin{equation}\label{eq:lowfrequencyresponse}
	\chi[\Omega\ll \Omega_m] \approx \frac{1}{m\Omega_m^2}\left(1+i\phi[\Omega] \right).
\end{equation}
Deviation from eq. \eqref{eq:lowfrequencyresponse} is observed in the measured response below 50 kHz.  The observed $1/f$ scaling is also present when the beam is absent, thus we attribute it to the photothermal response of the microdisk.  Also shown in \cref{fig:thermNoise}b is a multimode susceptibility model, including $n=3,5$ modes with frequencies $\Omega_\t{m}^{(n)}=n\cdot\Omega_\t{m}^{(1)}$ and assuming $m^{(n)}=m^{(1)}$ and $Q^{(n)}=Q^{(1)}$. Including these modes increases $|\chi[\Omega\ll\Omega_\t{m}]|$ by $50\%$, which is significantly larger than the estimated contribution to thermal noise ($<12\%$) because susceptibilities add incoherently in the latter case.

(3) Having provided evidence that the origin of the $1/f$ photocurrent noise is thermomechanical in origin and that the susceptibility of the nanobeam is that of a damped harmonic oscillator, we now compare the thermomechanical noise to the generalized model from the fluctuation dissipation theorem (\cref{SFth_general}a):
\begin{equation}\label{eq:Sxth2}
S_x^\t{th}[\Omega]= |\chi[\Omega]|^2\cdot4k_B T m\Omega_m\phi[\Omega],
\end{equation}
here ignoring radiation pressure damping (since it only affects the spectrum near resonance). Yellow and red curves in \cref{fig:thermNoise}a are models for velocity and structural damping, respectively, using $T=295$ K and $Q=2.3\cdot10^{5}$.  A clear deviation from the velocity-damping model is seen for $\Omega<\Omega_\t{m}$.  By contrast, the structural damping model matches the measured noise spectrum to within a factor of 2 over two decades in frequency.  This correspondence suggests that the internal loss angle is approximately constant over the same frequency range.

The frequency dependence of $\phi$ should in principle also be reflected in the $Q$ dispersion of higher order flexural modes, provided they are limited by the same internal loss mechanism. Naively, it is expected that $Q$ is mode-independent for a structurally damped resonator.  Due to stress, however, internal loss of each mode is diluted by a factor $D_\sigma^{(n)}$ depending on the mode shape and stress magnitude $\sigma$, resulting in an effective loss angle
\begin{equation}\label{eq:dilution}
	\phi^{(n)}[\Omega]=D_\sigma^{(n)}\cdot \phi_\t{intr}[\Omega]
\end{equation}   
where $\phi_\t{intr}[\Omega]$ is the intrinsic loss in the absence of stress \cite{gonzalez_brownian_1994,villanueva2014evidence,Kotthaus_damping_PRL_2010}.  In \cref{fig:phiOmega}, we use eq. \eqref{eq:dilution} to plot $\phi_\t{intr}[\Omega]$ from 10 kHz to 50 MHz.  Values for $\Omega\lesssim2\pi\cdot3.4$ MHz (blue line) are inferred from the broadband thermal noise spectrum by inverting eq. \eqref{eq:Sxth2}. Values for $\Omega>2\pi\cdot3.4$ MHz (green points) are obtained from the linewidth of thermal noise peaks of the $n=\{1,3,5,7,9,11,13\}$ out-of-plane flexural modes.  $D_\sigma^{(n)}$ is numerically calculated using a finite element model of the mode shapes \cite{Kotthaus_damping_PRL_2010}.  The dilution factor is a weak function of $n$ and is approximately $D_\sigma^{(1)}=1.2\cdot10^{-2}$ for the fundamental resonance. The inferred intrinsic loss angle $\phi_\t{intr}[\Omega<2\pi\cdot 3.4\,\t{MHz}]\approx 3\cdot 10^{-4}$ agrees well with a recent survey of stressed SiN thin film resonators which attributes a significant fraction of the loss to surface material \cite{villanueva2014evidence}.  Remarkably, we find that $\phi_\t{intr}$ remains constant to within a factor of 3 over the full four decade frequency window.  Similar broadband structural damping behavior has been observed in high-stress wires made of fused silica \cite{gonzalez_brownian_1995,kajima_wide-band_1999},
nylon \cite{bernardini_1999} and tungsten \cite{yang_2009}.


\section{Discussion of implications for broadband quantum optomechanics}\label{sec:discussion}

Thermomechanical noise fundamentally limits the performance of optomechanical transducers. We here consider two examples: quantum-enhanced force detection \cite{vyatchanin_variational_1995} and 
squeezed light generation \cite{fabre_quantum-noise_1994,mancini_quantum_1994}
using an optomechanical system.  Both of these applications can be discussed in terms of homodyne
measurement of the position of a mechancial oscillator. Such measurements are typically conducted in the 
unresolved-sideband regime, $\Omega_m\ll\kappa$, with the laser driving the cavity on resonance. We limit discussion to this case.

Under the aforement conditions, the detected homodyne spectrum, normalized to the shot noise level, can be expressed as
by \cite{sudhir_RT_QBA_2016},
\begin{equation}\label{eq:genHomSpectr}
\begin{split}
	S_I[\Omega]=1 &+\eta\frac{16 G^2 n_c}{\kappa}
	\left(S_x[\Omega]\sin^2(\theta) \frac{}{} \right. \\
	&\left. +\frac{\hbar}{2}\sin(2\theta)\mathrm{Re}\,\chi[\Omega]\right),
\end{split}
\end{equation}
where $\theta$ is the homodyne detection angle, $\eta$ is the optical detection efficiency and $n_c$ is the average number of photons in the optical cavity. 
The oscillator motion is driven by a thermal force (and zero-point fluctuations in quantum case), 
as well as the quantum back-action of the measurement (radiation pressure shot noise).  The resulting displacement noise spectrum is
\begin{equation}
\begin{split}
	S_{x}[\Omega]=|\chi[\Omega]|^2 4\hbar m\Omega_m^2 &\left[ \phi[\Omega] (n_\t{th}[\Omega]+\tfrac{1}{2})\frac{}{} \right.\\
		&\left. \frac{}{}+\phi[\Omega_m] n_\t{QBA}\right],
\end{split}
\end{equation}
where $n_\t{th}[\Omega]=(\exp(\hbar\Omega/k_B T)-1)^{-1}$ is the Bose thermal occupation of the oscillator and 
$n_\t{QBA}=\frac{4 G^2 x_\t{zpf}^2}{\kappa\Omega_m\phi[\Omega_m]}n_c$ is the phonon occupation due to measurement back-action (here $x_\t{zpf}=\sqrt{\hbar/2m\Omega_\t{m}}$ is the RMS zero-point displacement). In addition to 
increasing the oscillator occupation, measurement back-action creates correlations between quadratures of the meter optical field 
represented as the second term in brackets in the eq. \eqref{eq:genHomSpectr}. 
These correlations can be negative and, if the measurement is sufficiently strong, 
they can reduce the homodyne noise below the shot noise level (a signature of having created a squeezed state of light), or they can 
improve signal-to-noise ratio in detection of external force acting on the oscillator (an example of quantum-enhanced force sensing \cite{vyatchanin_variational_1995}). 
Due to the presence of thermal noise, both phenomena only become observable when the back-action force is comparable to the thermal force, 
i.e. $n_\t{QBA} \approx n_\t{th}[\Omega_\t{m}]$. 
Furthermore, correlations are limited to the vicinity of the mechanical resonance 
where the mechanical susceptibility is maximum. The experimentally demonstrated frequency ranges of generated 
ponderomotive squeezing $\Delta\Omega_\t{sq}$ have been so far at the percent level of the oscillator frequency:
$\Delta\Omega_\t{sq}/\Omega_\t{m}\approx1.3\times10^{-2}$~\cite{purdy_squeezing_2013}, $1.5\times10^{-1}$~\cite{safavi-naeini_squeezed_2013}, 
$7\times10^{-2}$~\cite{sudhir_appearance_2017}, and $3\times10^{-2}$~\cite{nielsen_multimode_2016}. 
Similar relative detuning ranges are considered in two recent experimental demonstrations of quantum-enhanced thermal force 
measurements: $\Delta\Omega/\Omega_\t{m}\approx 0.2\times 10^{-2}$~\cite{kampel_variational_2016} and
$2\times 10^{-1}$~\cite{sudhir_RT_QBA_2016}.

When restricting the frequency range of interest to around the mechanical resonance, dispersion of the loss angle can be safely 
neglected in most cases and the thermal noise be treated as white $n_\t{th}[\Omega]\approx n_\t{th}[\Omega_\t{m}]$. 
However, at large measurement strength, when $n_\t{QBA} \gtrsim n_\t{th}[\Omega_\t{m}]$, broadband optomechanical quantum 
correlations gradually become accessible and the full frequency dependence of thermal noise should be taken into account.
In the case of white thermal noise, the range over which squeezing can be observed increases as
\begin{equation}
	\frac{\Delta\Omega_\t{sq}}{\Gamma_\t{m}}\approx (n_\t{th}[\Omega_\t{m}]+n_{\mathrm{QBA}}),
\end{equation}    
and for the quantum-enhanced force sensitivity, 
\begin{equation}
	\frac{\Delta\Omega}{\Gamma_\t{m}}\approx \sqrt{\eta\, n_\t{QBA}(n_\t{th}[\Omega_\t{m}]+n_{\mathrm{QBA}})}.
\end{equation}
This imposes an additional requirement of $n_{\t{QBA}}>Q$ for observation of, e.g, squeezing at near zero frequency. 
This conclusion, however, is strongly altered if the oscillator dissipation has pronounced frequency dependence. 
In this case the maximum squeezing level is deteriorated by a factor
$(\Omega_\t{m}/\Omega)(\phi[\Omega]/\phi[\Omega_\t{m}])$ and the criterion for its 
observation at low frequencies ($\Omega\ll\Omega_\t{m}$) becomes 
\begin{equation}\label{eq:genDCQCorrCond}
	n_{\mathrm{QBA}}>\left\{\,Q,\,n_\t{th}[\Omega]\phi[\Omega]\,\right\}.
\end{equation}
This same condition 
holds for the ability to perform variational measurements at frequencies below mechanical resonance. In the case of structural 
damping it becomes increasingly demanding at lower frequencies since, $n_\t{th}\approx k_B T/\hbar\Omega$.

The consideration above applies to high-Q mechanical oscillators, which are currently the only practically relevant systems 
for optomechanics in the quantum regime. In the case of a viscously damped oscillator, the consequence of high-Q is that $\chi[\Omega]$ can be treated as purely real far below resonance ($\chi_{\mathrm{visc}}[\Omega\ll \Omega_\t{m}] \approx (1+i Q^{-1}(\Omega/\Omega_m))/m\Omega_\t{m}^2$). 
Given that it is the real part of the mechanical susceptibility that leads to ponderomotive squeezing via the correlation term
in \cref{eq:genHomSpectr}, the degree of squeezing generated by a viscously-damped oscillator is simply limited by 
the ratio $n_\t{QBA}/n_\t{th}$, and the detection efficiency $\eta$ alone. 
However, for a structurally damped oscillator, the mechanical susceptibility far below resonance $\chi_{\mathrm{struct}}[\Omega\ll \Omega_\t{m}] \approx (1+i Q^{-1})/m\Omega_\t{m}^2$, places another fundamental limit 
of $|\t{Im} (\chi_\t{m})/\chi_\t{m}|^2=Q^{-2}$ on the maximum degree of squeezing. Specifically,
this limits the achievable squeezing at DC, which has so far only been theoretically treated under the assumption of a viscously-damped
oscillator \cite{fabre_quantum-noise_1994,mancini_quantum_1994}. 

\section{Conclusion}

We have presented evidence that high-stress $\mathrm{Si_3 N_4}$ nanobeam resonators undergo dissipation that to a good precision can be described by a frequency-independent loss angle. Our measurements suggest that such ``structural damping" is an intrinsic property of $\mathrm{Si_3 N_4}$ thin films, supporting available data from literature on frequency scaling of the intrinsic quality factor. Correspondingly, the familiar assumption of a white thermal force is not applicable to $\mathrm{Si_3 N_4}$ nanomechanical resonators (e.g. nanobeams and nanomembranes) over a broad range of frequencies.  This may result in an underestimate of the thermal motion at low frequencies. Our results may also be viewed as a direct observation of a specific type of non-Markovian coupling between a nanomechanical
oscillator and its environment, complementing recent findings based on statistical analysis of near-resonant noise spectra \cite{groblacher_observation_2015}.
Finally, we discuss the implications of
structural damping for broadband operation of an optomechanical system in the quantum regime. 
We theoretically show the effect of structural damping is to limit the degree of quantum correlations of light
imparted by the optomechanical interaction at frequencies $\Omega\ll\Omega_\t{m}$.  Practically however, contemporary systems are far from reaching these limits.

\section*{Acknowledgements}

All samples were fabricated at the Center for MicroNanotechnology (CMi) at EPFL. Research was by an ERC advanced grant (QuREM), a Marie
Curie Initial Training Network in Cavity Quantum Optomechanics (CQOM), the Swiss National Science Foundation, and the NCCR of Quantum Engineering (QSIT). D.J.W. acknowledges support from the European Commission through a Marie Skłodowska-Curie Fellowship (IIF project 331985).

\section*{References}

\bibliography{bibliography,references}

\begin{thebibliography}{46}%
\makeatletter
\providecommand \@ifxundefined [1]{%
 \@ifx{#1\undefined}
}%
\providecommand \@ifnum [1]{%
 \ifnum #1\expandafter \@firstoftwo
 \else \expandafter \@secondoftwo
 \fi
}%
\providecommand \@ifx [1]{%
 \ifx #1\expandafter \@firstoftwo
 \else \expandafter \@secondoftwo
 \fi
}%
\providecommand \natexlab [1]{#1}%
\providecommand \enquote  [1]{``#1''}%
\providecommand \bibnamefont  [1]{#1}%
\providecommand \bibfnamefont [1]{#1}%
\providecommand \citenamefont [1]{#1}%
\providecommand \href@noop [0]{\@secondoftwo}%
\providecommand \href [0]{\begingroup \@sanitize@url \@href}%
\providecommand \@href[1]{\@@startlink{#1}\@@href}%
\providecommand \@@href[1]{\endgroup#1\@@endlink}%
\providecommand \@sanitize@url [0]{\catcode `\\12\catcode `\$12\catcode
  `\&12\catcode `\#12\catcode `\^12\catcode `\_12\catcode `\%12\relax}%
\providecommand \@@startlink[1]{}%
\providecommand \@@endlink[0]{}%
\providecommand \url  [0]{\begingroup\@sanitize@url \@url }%
\providecommand \@url [1]{\endgroup\@href {#1}{\urlprefix }}%
\providecommand \urlprefix  [0]{URL }%
\providecommand \Eprint [0]{\href }%
\providecommand \doibase [0]{http://dx.doi.org/}%
\providecommand \selectlanguage [0]{\@gobble}%
\providecommand \bibinfo  [0]{\@secondoftwo}%
\providecommand \bibfield  [0]{\@secondoftwo}%
\providecommand \translation [1]{[#1]}%
\providecommand \BibitemOpen [0]{}%
\providecommand \bibitemStop [0]{}%
\providecommand \bibitemNoStop [0]{.\EOS\space}%
\providecommand \EOS [0]{\spacefactor3000\relax}%
\providecommand \BibitemShut  [1]{\csname bibitem#1\endcsname}%
\let\auto@bib@innerbib\@empty
\bibitem [{\citenamefont {Binnig}\ \emph {et~al.}(1986)\citenamefont {Binnig},
  \citenamefont {Quate},\ and\ \citenamefont {Gerber}}]{binnig_1986}%
  \BibitemOpen
  \bibfield  {author} {\bibinfo {author} {\bibfnamefont {G.}~\bibnamefont
  {Binnig}}, \bibinfo {author} {\bibfnamefont {C.~F.}\ \bibnamefont {Quate}}, \
  and\ \bibinfo {author} {\bibfnamefont {C.}~\bibnamefont {Gerber}},\ }\href
  {http://journals.aps.org/prl/abstract/10.1103/PhysRevLett.56.930} {\bibfield
  {journal} {\bibinfo  {journal} {Physical Review Letters}\ }\textbf {\bibinfo
  {volume} {56}},\ \bibinfo {pages} {930} (\bibinfo {year} {1986})}\BibitemShut
  {NoStop}%
\bibitem [{\citenamefont {Rugar}\ \emph {et~al.}(1992)\citenamefont {Rugar},
  \citenamefont {Yannoni},\ and\ \citenamefont
  {Sidles}}]{rugar_mechanical_1992}%
  \BibitemOpen
  \bibfield  {author} {\bibinfo {author} {\bibfnamefont {D.}~\bibnamefont
  {Rugar}}, \bibinfo {author} {\bibfnamefont {C.~S.}\ \bibnamefont {Yannoni}},
  \ and\ \bibinfo {author} {\bibfnamefont {J.~A.}\ \bibnamefont {Sidles}},\
  }\href {\doibase 10.1038/360563a0} {\bibfield  {journal} {\bibinfo  {journal}
  {Nature}\ }\textbf {\bibinfo {volume} {360}},\ \bibinfo {pages} {563}
  (\bibinfo {year} {1992})}\BibitemShut {NoStop}%
\bibitem [{\citenamefont {Saulson}(1990)}]{saulson_thermal_1990}%
  \BibitemOpen
  \bibfield  {author} {\bibinfo {author} {\bibfnamefont {P.~R.}\ \bibnamefont
  {Saulson}},\ }\href
  {http://journals.aps.org/prd/abstract/10.1103/PhysRevD.42.2437} {\bibfield
  {journal} {\bibinfo  {journal} {Physical Review D}\ }\textbf {\bibinfo
  {volume} {42}},\ \bibinfo {pages} {2437} (\bibinfo {year}
  {1990})}\BibitemShut {NoStop}%
\bibitem [{spe()}]{spectraldensity}%
  \BibitemOpen
  \href@noop {} {\bibinfo  {journal} {In this paper we adopt the convention for
  a single-sided spectral density
  $S_x[\Omega]\equiv2\int_{-\infty}^{\infty}\langle x(t) x(t+\tau)\rangle
  e^{-i\Omega \tau}d\tau$ normalized such that $\langle x^2\rangle =
  \int_0^{\infty}S_x[\Omega]d\Omega/2\pi$}\ }\BibitemShut {NoStop}%
\bibitem [{\citenamefont {Imboden}\ and\ \citenamefont
  {Mohanty}(2014)}]{imboden2014dissipation}%
  \BibitemOpen
\bibfield  {journal} {  }\bibfield  {author} {\bibinfo {author} {\bibfnamefont
  {M.}~\bibnamefont {Imboden}}\ and\ \bibinfo {author} {\bibfnamefont
  {P.}~\bibnamefont {Mohanty}},\ }\href
  {http://www.sciencedirect.com/science/article/pii/S0370157313003475#}
  {\bibfield  {journal} {\bibinfo  {journal} {Phys. Rep.}\ }\textbf {\bibinfo
  {volume} {534}},\ \bibinfo {pages} {89} (\bibinfo {year} {2014})}\BibitemShut
  {NoStop}%
\bibitem [{\citenamefont {Aspelmeyer}\ \emph {et~al.}(2014)\citenamefont
  {Aspelmeyer}, \citenamefont {Kippenberg},\ and\ \citenamefont
  {Marquardt}}]{aspelmeyer_cavity_2015}%
  \BibitemOpen
  \bibfield  {author} {\bibinfo {author} {\bibfnamefont {M.}~\bibnamefont
  {Aspelmeyer}}, \bibinfo {author} {\bibfnamefont {T.~J.}\ \bibnamefont
  {Kippenberg}}, \ and\ \bibinfo {author} {\bibfnamefont {F.}~\bibnamefont
  {Marquardt}},\ }\href {\doibase 10.1103/RevModPhys.86.1391} {\bibfield
  {journal} {\bibinfo  {journal} {Rev. Mod. Phys.}\ }\textbf {\bibinfo {volume}
  {86}},\ \bibinfo {pages} {1391} (\bibinfo {year} {2014})}\BibitemShut
  {NoStop}%
\bibitem [{\citenamefont {Wilson}\ \emph {et~al.}(2015)\citenamefont {Wilson},
  \citenamefont {Sudhir}, \citenamefont {Piro}, \citenamefont {Schilling},
  \citenamefont {Ghadimi},\ and\ \citenamefont
  {Kippenberg}}]{wilson2015measurement}%
  \BibitemOpen
  \bibfield  {author} {\bibinfo {author} {\bibfnamefont {D.}~\bibnamefont
  {Wilson}}, \bibinfo {author} {\bibfnamefont {V.}~\bibnamefont {Sudhir}},
  \bibinfo {author} {\bibfnamefont {N.}~\bibnamefont {Piro}}, \bibinfo {author}
  {\bibfnamefont {R.}~\bibnamefont {Schilling}}, \bibinfo {author}
  {\bibfnamefont {A.}~\bibnamefont {Ghadimi}}, \ and\ \bibinfo {author}
  {\bibfnamefont {T.}~\bibnamefont {Kippenberg}},\ }\href
  {http://www.nature.com/nature/journal/v524/n7565/abs/nature14672.html}
  {\bibfield  {journal} {\bibinfo  {journal} {Nature}\ }\textbf {\bibinfo
  {volume} {524}},\ \bibinfo {pages} {325} (\bibinfo {year}
  {2015})}\BibitemShut {NoStop}%
\bibitem [{\citenamefont {Purdy}\ \emph
  {et~al.}(2013{\natexlab{a}})\citenamefont {Purdy}, \citenamefont {Peterson},\
  and\ \citenamefont {Regal}}]{purdy_observation_2013}%
  \BibitemOpen
  \bibfield  {author} {\bibinfo {author} {\bibfnamefont {T.~P.}\ \bibnamefont
  {Purdy}}, \bibinfo {author} {\bibfnamefont {R.~W.}\ \bibnamefont {Peterson}},
  \ and\ \bibinfo {author} {\bibfnamefont {C.~A.}\ \bibnamefont {Regal}},\
  }\href {http://www.sciencemag.org/cgi/doi/10.1126/science.1231282} {\bibfield
   {journal} {\bibinfo  {journal} {Science}\ }\textbf {\bibinfo {volume}
  {339}},\ \bibinfo {pages} {801} (\bibinfo {year}
  {2013}{\natexlab{a}})}\BibitemShut {NoStop}%
\bibitem [{\citenamefont {Teufel}\ \emph {et~al.}(2016)\citenamefont {Teufel},
  \citenamefont {Lecocq},\ and\ \citenamefont
  {Simmonds}}]{teufel2016overwhelming}%
  \BibitemOpen
  \bibfield  {author} {\bibinfo {author} {\bibfnamefont {J.}~\bibnamefont
  {Teufel}}, \bibinfo {author} {\bibfnamefont {F.}~\bibnamefont {Lecocq}}, \
  and\ \bibinfo {author} {\bibfnamefont {R.}~\bibnamefont {Simmonds}},\ }\href
  {http://journals.aps.org/prl/abstract/10.1103/PhysRevLett.116.013602}
  {\bibfield  {journal} {\bibinfo  {journal} {Phys. Rev. Lett.}\ }\textbf
  {\bibinfo {volume} {116}},\ \bibinfo {pages} {013602} (\bibinfo {year}
  {2016})}\BibitemShut {NoStop}%
\bibitem [{\citenamefont {Norte}\ \emph {et~al.}(2016)\citenamefont {Norte},
  \citenamefont {Moura},\ and\ \citenamefont
  {Gr{\"o}blacher}}]{norte2016mechanical}%
  \BibitemOpen
  \bibfield  {author} {\bibinfo {author} {\bibfnamefont {R.~A.}\ \bibnamefont
  {Norte}}, \bibinfo {author} {\bibfnamefont {J.~P.}\ \bibnamefont {Moura}}, \
  and\ \bibinfo {author} {\bibfnamefont {S.}~\bibnamefont {Gr{\"o}blacher}},\
  }\href {https://journals.aps.org/prl/abstract/10.1103/PhysRevLett.116.147202}
  {\bibfield  {journal} {\bibinfo  {journal} {Phys. Rev. Lett.}\ }\textbf
  {\bibinfo {volume} {116}},\ \bibinfo {pages} {147202} (\bibinfo {year}
  {2016})}\BibitemShut {NoStop}%
\bibitem [{\citenamefont {Reinhardt}\ \emph {et~al.}(2016)\citenamefont
  {Reinhardt}, \citenamefont {M{\"u}ller}, \citenamefont {Bourassa},\ and\
  \citenamefont {Sankey}}]{reinhardt2016ultralow}%
  \BibitemOpen
  \bibfield  {author} {\bibinfo {author} {\bibfnamefont {C.}~\bibnamefont
  {Reinhardt}}, \bibinfo {author} {\bibfnamefont {T.}~\bibnamefont
  {M{\"u}ller}}, \bibinfo {author} {\bibfnamefont {A.}~\bibnamefont
  {Bourassa}}, \ and\ \bibinfo {author} {\bibfnamefont {J.~C.}\ \bibnamefont
  {Sankey}},\ }\href
  {https://journals.aps.org/prx/abstract/10.1103/PhysRevX.6.021001} {\bibfield
  {journal} {\bibinfo  {journal} {Phys. Rev. X}\ }\textbf {\bibinfo {volume}
  {6}},\ \bibinfo {pages} {021001} (\bibinfo {year} {2016})}\BibitemShut
  {NoStop}%
\bibitem [{\citenamefont {Braginsky}\ and\ \citenamefont
  {Braginski{\u\i}}()}]{braginsky1995quantum}%
  \BibitemOpen
  \bibfield  {author} {\bibinfo {author} {\bibfnamefont {V.~B.}\ \bibnamefont
  {Braginsky}}\ and\ \bibinfo {author} {\bibnamefont {Braginski{\u\i}}},\
  }\href@noop {} {\emph {\bibinfo {title} {Quantum measurement}}}\BibitemShut
  {NoStop}%
\bibitem [{\citenamefont {Braginsky}\ and\ \citenamefont
  {Manukin}(1977)}]{braginsky_weak_1977}%
  \BibitemOpen
  \bibfield  {author} {\bibinfo {author} {\bibfnamefont {V.~B.}\ \bibnamefont
  {Braginsky}}\ and\ \bibinfo {author} {\bibfnamefont {A.~B.}\ \bibnamefont
  {Manukin}},\ }\href@noop {} {\emph {\bibinfo {title} {Measurement of Weak
  Forces in Physics Experiments}}}\ (\bibinfo  {publisher} {University of
  Chicago Press},\ \bibinfo {year} {1977})\BibitemShut {NoStop}%
\bibitem [{\citenamefont {Corbitt}\ and\ \citenamefont
  {Mavalvala}(2004)}]{corbitt2004review}%
  \BibitemOpen
  \bibfield  {author} {\bibinfo {author} {\bibfnamefont {T.}~\bibnamefont
  {Corbitt}}\ and\ \bibinfo {author} {\bibfnamefont {N.}~\bibnamefont
  {Mavalvala}},\ }\href
  {http://iopscience.iop.org/article/10.1088/1464-4266/6/8/008/meta} {\bibfield
   {journal} {\bibinfo  {journal} {Journal of Optics B: Quantum and
  Semiclassical Optics}\ }\textbf {\bibinfo {volume} {6}},\ \bibinfo {pages}
  {S675} (\bibinfo {year} {2004})}\BibitemShut {NoStop}%
\bibitem [{\citenamefont {Verbridge}\ \emph {et~al.}(2008)\citenamefont
  {Verbridge}, \citenamefont {Ilic}, \citenamefont {Craighead},\ and\
  \citenamefont {Parpia}}]{verbridge2008size}%
  \BibitemOpen
  \bibfield  {author} {\bibinfo {author} {\bibfnamefont {S.~S.}\ \bibnamefont
  {Verbridge}}, \bibinfo {author} {\bibfnamefont {R.}~\bibnamefont {Ilic}},
  \bibinfo {author} {\bibfnamefont {H.}~\bibnamefont {Craighead}}, \ and\
  \bibinfo {author} {\bibfnamefont {J.~M.}\ \bibnamefont {Parpia}},\ }\href
  {http://scitation.aip.org/content/aip/journal/apl/93/1/10.1063/1.2952762}
  {\bibfield  {journal} {\bibinfo  {journal} {App. Phys. Lett.}\ }\textbf
  {\bibinfo {volume} {93}},\ \bibinfo {pages} {013101} (\bibinfo {year}
  {2008})}\BibitemShut {NoStop}%
\bibitem [{\citenamefont {Sudhir}\ \emph
  {et~al.}(2017{\natexlab{a}})\citenamefont {Sudhir}, \citenamefont {Wilson},
  \citenamefont {Schilling}, \citenamefont {Sch\"utz}, \citenamefont {Fedorov},
  \citenamefont {Ghadimi}, \citenamefont {Nunnenkamp},\ and\ \citenamefont
  {Kippenberg}}]{sudhir2017appearance}%
  \BibitemOpen
  \bibfield  {author} {\bibinfo {author} {\bibfnamefont {V.}~\bibnamefont
  {Sudhir}}, \bibinfo {author} {\bibfnamefont {D.~J.}\ \bibnamefont {Wilson}},
  \bibinfo {author} {\bibfnamefont {R.}~\bibnamefont {Schilling}}, \bibinfo
  {author} {\bibfnamefont {H.}~\bibnamefont {Sch\"utz}}, \bibinfo {author}
  {\bibfnamefont {S.~A.}\ \bibnamefont {Fedorov}}, \bibinfo {author}
  {\bibfnamefont {A.~H.}\ \bibnamefont {Ghadimi}}, \bibinfo {author}
  {\bibfnamefont {A.}~\bibnamefont {Nunnenkamp}}, \ and\ \bibinfo {author}
  {\bibfnamefont {T.~J.}\ \bibnamefont {Kippenberg}},\ }\href {\doibase
  10.1103/PhysRevX.7.011001} {\bibfield  {journal} {\bibinfo  {journal} {Phys.
  Rev. X}\ }\textbf {\bibinfo {volume} {7}},\ \bibinfo {pages} {011001}
  (\bibinfo {year} {2017}{\natexlab{a}})}\BibitemShut {NoStop}%
\bibitem [{\citenamefont {Gonz{\'a}lez}\ and\ \citenamefont
  {Saulson}(1995)}]{gonzalez_brownian_1995}%
  \BibitemOpen
  \bibfield  {author} {\bibinfo {author} {\bibfnamefont {G.~I.}\ \bibnamefont
  {Gonz{\'a}lez}}\ and\ \bibinfo {author} {\bibfnamefont {P.~R.}\ \bibnamefont
  {Saulson}},\ }\href {\doibase 10.1016/0375-9601(95)00194-8} {\bibfield
  {journal} {\bibinfo  {journal} {Physics Letters A}\ }\textbf {\bibinfo
  {volume} {201}},\ \bibinfo {pages} {12} (\bibinfo {year} {1995})}\BibitemShut
  {NoStop}%
\bibitem [{\citenamefont {Bernardini}\ \emph {et~al.}(1999)\citenamefont
  {Bernardini}, \citenamefont {Majorana}, \citenamefont {Ogawa}, \citenamefont
  {Puppo}, \citenamefont {Rapagnini}, \citenamefont {Ricci},\ and\
  \citenamefont {Testi}}]{bernardini_1999}%
  \BibitemOpen
  \bibfield  {author} {\bibinfo {author} {\bibfnamefont {A.}~\bibnamefont
  {Bernardini}}, \bibinfo {author} {\bibfnamefont {E.}~\bibnamefont
  {Majorana}}, \bibinfo {author} {\bibfnamefont {Y.}~\bibnamefont {Ogawa}},
  \bibinfo {author} {\bibfnamefont {P.}~\bibnamefont {Puppo}}, \bibinfo
  {author} {\bibfnamefont {P.}~\bibnamefont {Rapagnini}}, \bibinfo {author}
  {\bibfnamefont {F.}~\bibnamefont {Ricci}}, \ and\ \bibinfo {author}
  {\bibfnamefont {G.}~\bibnamefont {Testi}},\ }\href
  {http://www.sciencedirect.com/science/article/pii/S0375960199001462}
  {\bibfield  {journal} {\bibinfo  {journal} {Physics Letters A}\ }\textbf
  {\bibinfo {volume} {255}},\ \bibinfo {pages} {142} (\bibinfo {year}
  {1999})}\BibitemShut {NoStop}%
\bibitem [{\citenamefont {Yang}\ \emph {et~al.}(2009)\citenamefont {Yang},
  \citenamefont {Tu}, \citenamefont {Shao}, \citenamefont {Li}, \citenamefont
  {Wang}, \citenamefont {Zhou},\ and\ \citenamefont {Luo}}]{yang_2009}%
  \BibitemOpen
  \bibfield  {author} {\bibinfo {author} {\bibfnamefont {S.}~\bibnamefont
  {Yang}}, \bibinfo {author} {\bibfnamefont {L.}~\bibnamefont {Tu}}, \bibinfo
  {author} {\bibfnamefont {C.}~\bibnamefont {Shao}}, \bibinfo {author}
  {\bibfnamefont {Q.}~\bibnamefont {Li}}, \bibinfo {author} {\bibfnamefont
  {Q.}~\bibnamefont {Wang}}, \bibinfo {author} {\bibfnamefont {Z.}~\bibnamefont
  {Zhou}}, \ and\ \bibinfo {author} {\bibfnamefont {J.}~\bibnamefont {Luo}},\
  }\href {https://journals.aps.org/prd/abstract/10.1103/PhysRevD.80.122005}
  {\bibfield  {journal} {\bibinfo  {journal} {Physical Review D}\ }\textbf
  {\bibinfo {volume} {80}},\ \bibinfo {pages} {122005} (\bibinfo {year}
  {2009})}\BibitemShut {NoStop}%
\bibitem [{\citenamefont {Neben}\ \emph {et~al.}(2012)\citenamefont {Neben},
  \citenamefont {Bodiya}, \citenamefont {Wipf}, \citenamefont {Oelker},
  \citenamefont {Corbitt},\ and\ \citenamefont
  {Mavalvala}}]{neben2012structural}%
  \BibitemOpen
  \bibfield  {author} {\bibinfo {author} {\bibfnamefont {A.~R.}\ \bibnamefont
  {Neben}}, \bibinfo {author} {\bibfnamefont {T.~P.}\ \bibnamefont {Bodiya}},
  \bibinfo {author} {\bibfnamefont {C.}~\bibnamefont {Wipf}}, \bibinfo {author}
  {\bibfnamefont {E.}~\bibnamefont {Oelker}}, \bibinfo {author} {\bibfnamefont
  {T.}~\bibnamefont {Corbitt}}, \ and\ \bibinfo {author} {\bibfnamefont
  {N.}~\bibnamefont {Mavalvala}},\ }\href
  {http://iopscience.iop.org/article/10.1088/1367-2630/14/11/115008/meta}
  {\bibfield  {journal} {\bibinfo  {journal} {New Journal of Physics}\ }\textbf
  {\bibinfo {volume} {14}},\ \bibinfo {pages} {115008} (\bibinfo {year}
  {2012})}\BibitemShut {NoStop}%
\bibitem [{\citenamefont {Villanueva}\ and\ \citenamefont
  {Schmid}(2014{\natexlab{a}})}]{villanueva_surface_loss_2014}%
  \BibitemOpen
  \bibfield  {author} {\bibinfo {author} {\bibfnamefont {L.}~\bibnamefont
  {Villanueva}}\ and\ \bibinfo {author} {\bibfnamefont {S.}~\bibnamefont
  {Schmid}},\ }\href {\doibase 10.1103/PhysRevLett.113.227201} {\bibfield
  {journal} {\bibinfo  {journal} {Physical Review Letters}\ }\textbf {\bibinfo
  {volume} {113}},\ \bibinfo {pages} {227201} (\bibinfo {year}
  {2014}{\natexlab{a}})}\BibitemShut {NoStop}%
\bibitem [{\citenamefont {Gonz{\'a}lez}\ and\ \citenamefont
  {Saulson}(1994{\natexlab{a}})}]{gonzalez1994brownian}%
  \BibitemOpen
  \bibfield  {author} {\bibinfo {author} {\bibfnamefont {G.~I.}\ \bibnamefont
  {Gonz{\'a}lez}}\ and\ \bibinfo {author} {\bibfnamefont {P.~R.}\ \bibnamefont
  {Saulson}},\ }\href
  {http://scitation.aip.org/content/asa/journal/jasa/96/1/10.1121/1.410467}
  {\bibfield  {journal} {\bibinfo  {journal} {J. Acoust. Soc. Am.}\ }\textbf
  {\bibinfo {volume} {96}},\ \bibinfo {pages} {207} (\bibinfo {year}
  {1994}{\natexlab{a}})}\BibitemShut {NoStop}%
\bibitem [{\citenamefont {Unterreithmeier}\ \emph {et~al.}(2010)\citenamefont
  {Unterreithmeier}, \citenamefont {Faust},\ and\ \citenamefont
  {Kotthaus}}]{Kotthaus_damping_PRL_2010}%
  \BibitemOpen
  \bibfield  {author} {\bibinfo {author} {\bibfnamefont {Q.~P.}\ \bibnamefont
  {Unterreithmeier}}, \bibinfo {author} {\bibfnamefont {T.}~\bibnamefont
  {Faust}}, \ and\ \bibinfo {author} {\bibfnamefont {J.~P.}\ \bibnamefont
  {Kotthaus}},\ }\href {\doibase 10.1103/PhysRevLett.105.027205} {\bibfield
  {journal} {\bibinfo  {journal} {Physical Review Letters}\ }\textbf {\bibinfo
  {volume} {105}},\ \bibinfo {pages} {027205} (\bibinfo {year}
  {2010})}\BibitemShut {NoStop}%
\bibitem [{\citenamefont {Chakram}\ \emph {et~al.}(2014)\citenamefont
  {Chakram}, \citenamefont {Patil}, \citenamefont {Chang},\ and\ \citenamefont
  {Vengalattore}}]{chakram_dissipation_2014}%
  \BibitemOpen
  \bibfield  {author} {\bibinfo {author} {\bibfnamefont {S.}~\bibnamefont
  {Chakram}}, \bibinfo {author} {\bibfnamefont {Y.}~\bibnamefont {Patil}},
  \bibinfo {author} {\bibfnamefont {L.}~\bibnamefont {Chang}}, \ and\ \bibinfo
  {author} {\bibfnamefont {M.}~\bibnamefont {Vengalattore}},\ }\href {\doibase
  10.1103/PhysRevLett.112.127201} {\bibfield  {journal} {\bibinfo  {journal}
  {Physical Review Letters}\ }\textbf {\bibinfo {volume} {112}},\ \bibinfo
  {pages} {127201} (\bibinfo {year} {2014})}\BibitemShut {NoStop}%
\bibitem [{\citenamefont {Yu}\ \emph {et~al.}(2012)\citenamefont {Yu},
  \citenamefont {Purdy},\ and\ \citenamefont
  {Regal}}]{yuRegal_control_damping_2012}%
  \BibitemOpen
  \bibfield  {author} {\bibinfo {author} {\bibfnamefont {P.-L.}\ \bibnamefont
  {Yu}}, \bibinfo {author} {\bibfnamefont {T.~P.}\ \bibnamefont {Purdy}}, \
  and\ \bibinfo {author} {\bibfnamefont {C.~A.}\ \bibnamefont {Regal}},\ }\href
  {\doibase 10.1103/PhysRevLett.108.083603} {\bibfield  {journal} {\bibinfo
  {journal} {Physical Review Letters}\ }\textbf {\bibinfo {volume} {108}},\
  \bibinfo {pages} {083603} (\bibinfo {year} {2012})}\BibitemShut {NoStop}%
\bibitem [{\citenamefont {Ghadimi}\ \emph {et~al.}(2016)\citenamefont
  {Ghadimi}, \citenamefont {Wilson},\ and\ \citenamefont
  {Kippenberg}}]{ghadimi_dissipation_2016}%
  \BibitemOpen
  \bibfield  {author} {\bibinfo {author} {\bibfnamefont {A.~H.}\ \bibnamefont
  {Ghadimi}}, \bibinfo {author} {\bibfnamefont {D.~J.}\ \bibnamefont {Wilson}},
  \ and\ \bibinfo {author} {\bibfnamefont {T.~J.}\ \bibnamefont {Kippenberg}},\
  }\href {http://arxiv.org/abs/1603.01605} {\bibfield  {journal} {\bibinfo
  {journal} {arXiv:1603.01605}\ } (\bibinfo {year} {2016})}\BibitemShut
  {NoStop}%
\bibitem [{\citenamefont {Martin}\ \emph {et~al.}(2008)\citenamefont {Martin},
  \citenamefont {Houston}, \citenamefont {Baldwin},\ and\ \citenamefont
  {Zalalutdinov}}]{martin_gas_damp_2008}%
  \BibitemOpen
  \bibfield  {author} {\bibinfo {author} {\bibfnamefont {M.~J.}\ \bibnamefont
  {Martin}}, \bibinfo {author} {\bibfnamefont {B.~H.}\ \bibnamefont {Houston}},
  \bibinfo {author} {\bibfnamefont {J.~W.}\ \bibnamefont {Baldwin}}, \ and\
  \bibinfo {author} {\bibfnamefont {M.~K.}\ \bibnamefont {Zalalutdinov}},\
  }\href {\doibase 10.1109/JMEMS.2008.916321} {\bibfield  {journal} {\bibinfo
  {journal} {Journal of Microelectromechanical Systems}\ }\textbf {\bibinfo
  {volume} {17}},\ \bibinfo {pages} {503} (\bibinfo {year} {2008})}\BibitemShut
  {NoStop}%
\bibitem [{\citenamefont {Bao}\ and\ \citenamefont
  {Yang}(2007)}]{bao_squeeze_2007}%
  \BibitemOpen
  \bibfield  {author} {\bibinfo {author} {\bibfnamefont {M.}~\bibnamefont
  {Bao}}\ and\ \bibinfo {author} {\bibfnamefont {H.}~\bibnamefont {Yang}},\
  }\href {\doibase 10.1016/j.sna.2007.01.008} {\bibfield  {journal} {\bibinfo
  {journal} {Sensors and Actuators A}\ }\textbf {\bibinfo {volume} {136}},\
  \bibinfo {pages} {3} (\bibinfo {year} {2007})}\BibitemShut {NoStop}%
\bibitem [{\citenamefont {Anetsberger}\ \emph {et~al.}(2009)\citenamefont
  {Anetsberger}, \citenamefont {Arcizet}, \citenamefont {Unterreithmeier},
  \citenamefont {Rivi{\`e}re}, \citenamefont {Schliesser}, \citenamefont
  {Weig}, \citenamefont {Kotthaus},\ and\ \citenamefont
  {Kippenberg}}]{anetsberger_near-field_2009}%
  \BibitemOpen
  \bibfield  {author} {\bibinfo {author} {\bibfnamefont {G.}~\bibnamefont
  {Anetsberger}}, \bibinfo {author} {\bibfnamefont {O.}~\bibnamefont
  {Arcizet}}, \bibinfo {author} {\bibfnamefont {Q.~P.}\ \bibnamefont
  {Unterreithmeier}}, \bibinfo {author} {\bibfnamefont {R.}~\bibnamefont
  {Rivi{\`e}re}}, \bibinfo {author} {\bibfnamefont {A.}~\bibnamefont
  {Schliesser}}, \bibinfo {author} {\bibfnamefont {E.~M.}\ \bibnamefont
  {Weig}}, \bibinfo {author} {\bibfnamefont {J.~P.}\ \bibnamefont {Kotthaus}},
  \ and\ \bibinfo {author} {\bibfnamefont {T.~J.}\ \bibnamefont {Kippenberg}},\
  }\href {http://www.nature.com/doifinder/10.1038/nphys1425} {\bibfield
  {journal} {\bibinfo  {journal} {Nat. Phys.}\ }\textbf {\bibinfo {volume}
  {5}},\ \bibinfo {pages} {909} (\bibinfo {year} {2009})}\BibitemShut {NoStop}%
\bibitem [{\citenamefont {Schilling}\ \emph {et~al.}(2016)\citenamefont
  {Schilling}, \citenamefont {Sch{\"u}tz}, \citenamefont {Ghadimi},
  \citenamefont {Sudhir}, \citenamefont {Wilson},\ and\ \citenamefont
  {Kippenberg}}]{schilling_near-field_2016}%
  \BibitemOpen
  \bibfield  {author} {\bibinfo {author} {\bibfnamefont {R.}~\bibnamefont
  {Schilling}}, \bibinfo {author} {\bibfnamefont {H.}~\bibnamefont
  {Sch{\"u}tz}}, \bibinfo {author} {\bibfnamefont {A.}~\bibnamefont {Ghadimi}},
  \bibinfo {author} {\bibfnamefont {V.}~\bibnamefont {Sudhir}}, \bibinfo
  {author} {\bibfnamefont {D.}~\bibnamefont {Wilson}}, \ and\ \bibinfo {author}
  {\bibfnamefont {T.}~\bibnamefont {Kippenberg}},\ }\href {\doibase
  10.1103/PhysRevApplied.5.054019} {\bibfield  {journal} {\bibinfo  {journal}
  {Physical Review Applied}\ }\textbf {\bibinfo {volume} {5}},\ \bibinfo
  {pages} {054019} (\bibinfo {year} {2016})}\BibitemShut {NoStop}%
\bibitem [{\citenamefont {Gorodetsky}\ \emph {et~al.}(2010)\citenamefont
  {Gorodetsky}, \citenamefont {Schliesser}, \citenamefont {Anetsberger},
  \citenamefont {Deleglise},\ and\ \citenamefont
  {Kippenberg}}]{gorodetsky_determination_2010}%
  \BibitemOpen
  \bibfield  {author} {\bibinfo {author} {\bibfnamefont {M.~L.}\ \bibnamefont
  {Gorodetsky}}, \bibinfo {author} {\bibfnamefont {A.}~\bibnamefont
  {Schliesser}}, \bibinfo {author} {\bibfnamefont {G.}~\bibnamefont
  {Anetsberger}}, \bibinfo {author} {\bibfnamefont {S.}~\bibnamefont
  {Deleglise}}, \ and\ \bibinfo {author} {\bibfnamefont {T.~J.}\ \bibnamefont
  {Kippenberg}},\ }\href
  {http://www.opticsinfobase.org/oe/fulltext.cfm?uri=oe-18-22-23236} {\bibfield
   {journal} {\bibinfo  {journal} {Opt. Exp.}\ }\textbf {\bibinfo {volume}
  {18}},\ \bibinfo {pages} {23236} (\bibinfo {year} {2010})}\BibitemShut
  {NoStop}%
\bibitem [{\citenamefont {Sudhir}\ \emph {et~al.}(2016)\citenamefont {Sudhir},
  \citenamefont {Schilling}, \citenamefont {Fedorov}, \citenamefont {Schuetz},
  \citenamefont {Wilson},\ and\ \citenamefont
  {Kippenberg}}]{sudhir_RT_QBA_2016}%
  \BibitemOpen
  \bibfield  {author} {\bibinfo {author} {\bibfnamefont {V.}~\bibnamefont
  {Sudhir}}, \bibinfo {author} {\bibfnamefont {R.}~\bibnamefont {Schilling}},
  \bibinfo {author} {\bibfnamefont {S.~A.}\ \bibnamefont {Fedorov}}, \bibinfo
  {author} {\bibfnamefont {H.}~\bibnamefont {Schuetz}}, \bibinfo {author}
  {\bibfnamefont {D.~J.}\ \bibnamefont {Wilson}}, \ and\ \bibinfo {author}
  {\bibfnamefont {T.~J.}\ \bibnamefont {Kippenberg}},\ }\href
  {http://arxiv.org/abs/1608.00699} {\bibfield  {journal} {\bibinfo  {journal}
  {arXiv:1608.00699}\ } (\bibinfo {year} {2016})},\ \Eprint
  {http://arxiv.org/abs/1608.00699} {1608.00699} \BibitemShut {NoStop}%
\bibitem [{\citenamefont {Sudhir}\ \emph
  {et~al.}(2017{\natexlab{b}})\citenamefont {Sudhir}, \citenamefont {Wilson},
  \citenamefont {Schilling}, \citenamefont {Sch{\"u}tz}, \citenamefont
  {Fedorov}, \citenamefont {Ghadimi}, \citenamefont {Nunnenkamp},\ and\
  \citenamefont {Kippenberg}}]{sudhir_appearance_2017}%
  \BibitemOpen
  \bibfield  {author} {\bibinfo {author} {\bibfnamefont {V.}~\bibnamefont
  {Sudhir}}, \bibinfo {author} {\bibfnamefont {D.}~\bibnamefont {Wilson}},
  \bibinfo {author} {\bibfnamefont {R.}~\bibnamefont {Schilling}}, \bibinfo
  {author} {\bibfnamefont {H.}~\bibnamefont {Sch{\"u}tz}}, \bibinfo {author}
  {\bibfnamefont {S.}~\bibnamefont {Fedorov}}, \bibinfo {author} {\bibfnamefont
  {A.}~\bibnamefont {Ghadimi}}, \bibinfo {author} {\bibfnamefont
  {A.}~\bibnamefont {Nunnenkamp}}, \ and\ \bibinfo {author} {\bibfnamefont
  {T.}~\bibnamefont {Kippenberg}},\ }\href {\doibase 10.1103/PhysRevX.7.011001}
  {\bibfield  {journal} {\bibinfo  {journal} {Physical Review X}\ }\textbf
  {\bibinfo {volume} {7}},\ \bibinfo {pages} {011001} (\bibinfo {year}
  {2017}{\natexlab{b}})}\BibitemShut {NoStop}%
\bibitem [{\citenamefont {Gorodetsky}\ and\ \citenamefont
  {Grudinin}(2004)}]{gorodetsky2004fundamental}%
  \BibitemOpen
  \bibfield  {author} {\bibinfo {author} {\bibfnamefont {M.~L.}\ \bibnamefont
  {Gorodetsky}}\ and\ \bibinfo {author} {\bibfnamefont {I.~S.}\ \bibnamefont
  {Grudinin}},\ }\href@noop {} {\bibfield  {journal} {\bibinfo  {journal} {JOSA
  B}\ }\textbf {\bibinfo {volume} {21}},\ \bibinfo {pages} {697} (\bibinfo
  {year} {2004})}\BibitemShut {NoStop}%
\bibitem [{\citenamefont {Anetsberger}\ \emph {et~al.}(2010)\citenamefont
  {Anetsberger}, \citenamefont {Gavartin}, \citenamefont {Arcizet},
  \citenamefont {Unterreithmeier}, \citenamefont {Weig}, \citenamefont
  {Gorodetsky}, \citenamefont {Kotthaus},\ and\ \citenamefont
  {Kippenberg}}]{anetsberger_measuring_2010}%
  \BibitemOpen
  \bibfield  {author} {\bibinfo {author} {\bibfnamefont {G.}~\bibnamefont
  {Anetsberger}}, \bibinfo {author} {\bibfnamefont {E.}~\bibnamefont
  {Gavartin}}, \bibinfo {author} {\bibfnamefont {O.}~\bibnamefont {Arcizet}},
  \bibinfo {author} {\bibfnamefont {Q.~P.}\ \bibnamefont {Unterreithmeier}},
  \bibinfo {author} {\bibfnamefont {E.~M.}\ \bibnamefont {Weig}}, \bibinfo
  {author} {\bibfnamefont {M.~L.}\ \bibnamefont {Gorodetsky}}, \bibinfo
  {author} {\bibfnamefont {J.~P.}\ \bibnamefont {Kotthaus}}, \ and\ \bibinfo
  {author} {\bibfnamefont {T.~J.}\ \bibnamefont {Kippenberg}},\ }\href
  {http://link.aps.org/doi/10.1103/PhysRevA.82.061804} {\bibfield  {journal}
  {\bibinfo  {journal} {Phys. Rev. A}\ }\textbf {\bibinfo {volume} {82}}
  (\bibinfo {year} {2010})}\BibitemShut {NoStop}%
\bibitem [{\citenamefont {Gonz{\'a}lez}\ and\ \citenamefont
  {Saulson}(1994{\natexlab{b}})}]{gonzalez_brownian_1994}%
  \BibitemOpen
  \bibfield  {author} {\bibinfo {author} {\bibfnamefont {G.~I.}\ \bibnamefont
  {Gonz{\'a}lez}}\ and\ \bibinfo {author} {\bibfnamefont {P.~R.}\ \bibnamefont
  {Saulson}},\ }\href {\doibase 10.1121/1.410467} {\bibfield  {journal}
  {\bibinfo  {journal} {The Journal of the Acoustical Society of America}\
  }\textbf {\bibinfo {volume} {96}},\ \bibinfo {pages} {207} (\bibinfo {year}
  {1994}{\natexlab{b}})}\BibitemShut {NoStop}%
\bibitem [{\citenamefont {Villanueva}\ and\ \citenamefont
  {Schmid}(2014{\natexlab{b}})}]{villanueva2014evidence}%
  \BibitemOpen
  \bibfield  {author} {\bibinfo {author} {\bibfnamefont {L.~G.}\ \bibnamefont
  {Villanueva}}\ and\ \bibinfo {author} {\bibfnamefont {S.}~\bibnamefont
  {Schmid}},\ }\href
  {http://journals.aps.org/prl/abstract/10.1103/PhysRevLett.113.227201}
  {\bibfield  {journal} {\bibinfo  {journal} {Phys. Rev. Lett.}\ }\textbf
  {\bibinfo {volume} {113}},\ \bibinfo {pages} {227201} (\bibinfo {year}
  {2014}{\natexlab{b}})}\BibitemShut {NoStop}%
\bibitem [{\citenamefont {Kajima}\ \emph {et~al.}(1999)\citenamefont {Kajima},
  \citenamefont {Kusumi}, \citenamefont {Moriwaki},\ and\ \citenamefont
  {Mio}}]{kajima_wide-band_1999}%
  \BibitemOpen
  \bibfield  {author} {\bibinfo {author} {\bibfnamefont {M.}~\bibnamefont
  {Kajima}}, \bibinfo {author} {\bibfnamefont {N.}~\bibnamefont {Kusumi}},
  \bibinfo {author} {\bibfnamefont {S.}~\bibnamefont {Moriwaki}}, \ and\
  \bibinfo {author} {\bibfnamefont {N.}~\bibnamefont {Mio}},\ }\href {\doibase
  10.1016/S0375-9601(99)00828-2} {\bibfield  {journal} {\bibinfo  {journal}
  {Physics Letters A}\ }\textbf {\bibinfo {volume} {264}},\ \bibinfo {pages}
  {251} (\bibinfo {year} {1999})}\BibitemShut {NoStop}%
\bibitem [{\citenamefont {Vyatchanin}\ and\ \citenamefont
  {Zubova}(1995)}]{vyatchanin_variational_1995}%
  \BibitemOpen
  \bibfield  {author} {\bibinfo {author} {\bibfnamefont {S.~P.}\ \bibnamefont
  {Vyatchanin}}\ and\ \bibinfo {author} {\bibfnamefont {E.~A.}\ \bibnamefont
  {Zubova}},\ }\href {\doibase 10.1016/0375-9601(95)00280-G} {\bibfield
  {journal} {\bibinfo  {journal} {Physics Letters A}\ }\textbf {\bibinfo
  {volume} {201}},\ \bibinfo {pages} {269} (\bibinfo {year}
  {1995})}\BibitemShut {NoStop}%
\bibitem [{\citenamefont {Fabre}\ \emph {et~al.}(1994)\citenamefont {Fabre},
  \citenamefont {Pinard}, \citenamefont {Bourzeix}, \citenamefont {Heidmann},
  \citenamefont {Giacobino},\ and\ \citenamefont
  {Reynaud}}]{fabre_quantum-noise_1994}%
  \BibitemOpen
  \bibfield  {author} {\bibinfo {author} {\bibfnamefont {C.}~\bibnamefont
  {Fabre}}, \bibinfo {author} {\bibfnamefont {M.}~\bibnamefont {Pinard}},
  \bibinfo {author} {\bibfnamefont {S.}~\bibnamefont {Bourzeix}}, \bibinfo
  {author} {\bibfnamefont {A.}~\bibnamefont {Heidmann}}, \bibinfo {author}
  {\bibfnamefont {E.}~\bibnamefont {Giacobino}}, \ and\ \bibinfo {author}
  {\bibfnamefont {S.}~\bibnamefont {Reynaud}},\ }\href {\doibase
  10.1103/PhysRevA.49.1337} {\bibfield  {journal} {\bibinfo  {journal} {Phys.
  Rev. A}\ }\textbf {\bibinfo {volume} {49}},\ \bibinfo {pages} {1337}
  (\bibinfo {year} {1994})}\BibitemShut {NoStop}%
\bibitem [{\citenamefont {Mancini}\ and\ \citenamefont
  {Tombesi}(1994)}]{mancini_quantum_1994}%
  \BibitemOpen
  \bibfield  {author} {\bibinfo {author} {\bibfnamefont {S.}~\bibnamefont
  {Mancini}}\ and\ \bibinfo {author} {\bibfnamefont {P.}~\bibnamefont
  {Tombesi}},\ }\href
  {http://journals.aps.org/pra/abstract/10.1103/PhysRevA.49.4055} {\bibfield
  {journal} {\bibinfo  {journal} {Physical Review A}\ }\textbf {\bibinfo
  {volume} {49}},\ \bibinfo {pages} {4055} (\bibinfo {year}
  {1994})}\BibitemShut {NoStop}%
\bibitem [{\citenamefont {Purdy}\ \emph
  {et~al.}(2013{\natexlab{b}})\citenamefont {Purdy}, \citenamefont {Yu},
  \citenamefont {Peterson}, \citenamefont {Kampel},\ and\ \citenamefont
  {Regal}}]{purdy_squeezing_2013}%
  \BibitemOpen
  \bibfield  {author} {\bibinfo {author} {\bibfnamefont {T.~P.}\ \bibnamefont
  {Purdy}}, \bibinfo {author} {\bibfnamefont {P.-L.}\ \bibnamefont {Yu}},
  \bibinfo {author} {\bibfnamefont {R.~W.}\ \bibnamefont {Peterson}}, \bibinfo
  {author} {\bibfnamefont {N.~S.}\ \bibnamefont {Kampel}}, \ and\ \bibinfo
  {author} {\bibfnamefont {C.~A.}\ \bibnamefont {Regal}},\ }\href {\doibase
  10.1103/PhysRevX.3.031012} {\bibfield  {journal} {\bibinfo  {journal}
  {Physical Review X}\ }\textbf {\bibinfo {volume} {3}} (\bibinfo {year}
  {2013}{\natexlab{b}}),\ 10.1103/PhysRevX.3.031012}\BibitemShut {NoStop}%
\bibitem [{\citenamefont {Safavi-Naeini}\ \emph {et~al.}(2013)\citenamefont
  {Safavi-Naeini}, \citenamefont {Gr{\"o}blacher}, \citenamefont {Hill},
  \citenamefont {Chan}, \citenamefont {Aspelmeyer},\ and\ \citenamefont
  {Painter}}]{safavi-naeini_squeezed_2013}%
  \BibitemOpen
  \bibfield  {author} {\bibinfo {author} {\bibfnamefont {A.~H.}\ \bibnamefont
  {Safavi-Naeini}}, \bibinfo {author} {\bibfnamefont {S.}~\bibnamefont
  {Gr{\"o}blacher}}, \bibinfo {author} {\bibfnamefont {J.~T.}\ \bibnamefont
  {Hill}}, \bibinfo {author} {\bibfnamefont {J.}~\bibnamefont {Chan}}, \bibinfo
  {author} {\bibfnamefont {M.}~\bibnamefont {Aspelmeyer}}, \ and\ \bibinfo
  {author} {\bibfnamefont {O.}~\bibnamefont {Painter}},\ }\href {\doibase
  10.1038/nature12307} {\bibfield  {journal} {\bibinfo  {journal} {Nature}\
  }\textbf {\bibinfo {volume} {500}},\ \bibinfo {pages} {185} (\bibinfo {year}
  {2013})}\BibitemShut {NoStop}%
\bibitem [{\citenamefont {Nielsen}\ \emph {et~al.}(2017)\citenamefont
  {Nielsen}, \citenamefont {Tsaturyan}, \citenamefont {M{\o}ller},
  \citenamefont {Polzik},\ and\ \citenamefont
  {Schliesser}}]{nielsen_multimode_2016}%
  \BibitemOpen
  \bibfield  {author} {\bibinfo {author} {\bibfnamefont {W.~H.~P.}\
  \bibnamefont {Nielsen}}, \bibinfo {author} {\bibfnamefont {Y.}~\bibnamefont
  {Tsaturyan}}, \bibinfo {author} {\bibfnamefont {C.~B.}\ \bibnamefont
  {M{\o}ller}}, \bibinfo {author} {\bibfnamefont {E.~S.}\ \bibnamefont
  {Polzik}}, \ and\ \bibinfo {author} {\bibfnamefont {A.}~\bibnamefont
  {Schliesser}},\ }\href {http://www.pnas.org/content/114/1/62} {\bibfield
  {journal} {\bibinfo  {journal} {Proceedings of the National Academy of
  Sciences}\ }\textbf {\bibinfo {volume} {114}},\ \bibinfo {pages} {62}
  (\bibinfo {year} {2017})}\BibitemShut {NoStop}%
\bibitem [{\citenamefont {Kampel}\ \emph {et~al.}(2016)\citenamefont {Kampel},
  \citenamefont {Peterson}, \citenamefont {Fischer}, \citenamefont {Yu},
  \citenamefont {Cicak}, \citenamefont {Simmonds}, \citenamefont {Lehnert},\
  and\ \citenamefont {Regal}}]{kampel_variational_2016}%
  \BibitemOpen
  \bibfield  {author} {\bibinfo {author} {\bibfnamefont {N.~S.}\ \bibnamefont
  {Kampel}}, \bibinfo {author} {\bibfnamefont {R.~W.}\ \bibnamefont
  {Peterson}}, \bibinfo {author} {\bibfnamefont {R.}~\bibnamefont {Fischer}},
  \bibinfo {author} {\bibfnamefont {P.-L.}\ \bibnamefont {Yu}}, \bibinfo
  {author} {\bibfnamefont {K.}~\bibnamefont {Cicak}}, \bibinfo {author}
  {\bibfnamefont {R.~W.}\ \bibnamefont {Simmonds}}, \bibinfo {author}
  {\bibfnamefont {K.~W.}\ \bibnamefont {Lehnert}}, \ and\ \bibinfo {author}
  {\bibfnamefont {C.~A.}\ \bibnamefont {Regal}},\ }\href
  {http://arxiv.org/abs/1607.06831} {\bibfield  {journal} {\bibinfo  {journal}
  {arXiv:1607.06831}\ } (\bibinfo {year} {2016})}\BibitemShut {NoStop}%
\bibitem [{\citenamefont {Gr{\"o}blacher}\ \emph {et~al.}(2015)\citenamefont
  {Gr{\"o}blacher}, \citenamefont {Trubarov}, \citenamefont {Prigge},
  \citenamefont {Cole}, \citenamefont {Aspelmeyer},\ and\ \citenamefont
  {Eisert}}]{groblacher_observation_2015}%
  \BibitemOpen
  \bibfield  {author} {\bibinfo {author} {\bibfnamefont {S.}~\bibnamefont
  {Gr{\"o}blacher}}, \bibinfo {author} {\bibfnamefont {A.}~\bibnamefont
  {Trubarov}}, \bibinfo {author} {\bibfnamefont {N.}~\bibnamefont {Prigge}},
  \bibinfo {author} {\bibfnamefont {G.~D.}\ \bibnamefont {Cole}}, \bibinfo
  {author} {\bibfnamefont {M.}~\bibnamefont {Aspelmeyer}}, \ and\ \bibinfo
  {author} {\bibfnamefont {J.}~\bibnamefont {Eisert}},\ }\href {\doibase
  10.1038/ncomms8606} {\bibfield  {journal} {\bibinfo  {journal} {Nature
  Communications}\ }\textbf {\bibinfo {volume} {6}},\ \bibinfo {pages} {7606}
  (\bibinfo {year} {2015})}\BibitemShut {NoStop}%
\end{thebibliography}%

\end{document}